\documentclass[12pt]{iopart}

\usepackage{graphicx}

\begin{document}

\title[Morphological instabilities of stratified epithelia]{Morphological instabilities of stratified epithelia: a mechanical instability in tumour formation}

\author{Thomas Risler$^{1,2,3}$ and Markus Basan$^{1,2,3}$\footnote{Present address: Center for Theoretical Biological Physics -- UC San Diego, MC 0374 -- 9500 Gilman Drive -- La Jolla, CA 92093-0374}}
\address{$^1$ Institut Curie, Centre de Recherche, UMR 168, 26 rue d'Ulm, F-75005, Paris, France}
\address{$^2$ UPMC Univ Paris 06, UMR 168, F-75005, Paris, France}
\address{$^3$ CNRS, UMR 168, F-75005, Paris, France}
\ead{thomas.risler@curie.fr}

\begin{abstract}
Interfaces between stratified epithelia and their supporting stromas commonly exhibit irregular shapes. Undulations are particularly pronounced in dysplastic tissues and typically evolve into long, finger-like protrusions in carcinomas. In a previous work [Basan {\it et al.}, {\it Phys. Rev. Lett.} {\bf 106} 158101, 2011], we demonstrated that an instability arising from viscous shear stresses caused by the constant flow due to cell turnover in the epithelium could drive this phenomenon. While interfacial tension between the two tissues as well as mechanical resistance of the stroma tend to maintain a flat interface, an instability occurs for sufficiently large viscosity, cell-division rate and thickness of the dividing region in the epithelium. Here, extensions of this work are presented, where cell division in the epithelium is coupled to the local concentration of nutrients or growth factors diffusing from the stroma. This enhances the instability by a mechanism similar to that of the Mullins-Sekerka instability in single-diffusion processes of crystal growth. We furthermore present the instability for the generalized case of a viscoelastic stroma.
\end{abstract}


\submitto{\NJP}

\maketitle

\section{Introduction}

Most human cancers are carcinomas, tumours that originate in epithelial tissues~\cite{weinberg2007bc}. From there, they may invade through the basement membrane into the supporting connective tissue---the stroma---and eventually lead to the formation of metastases~\cite{Nguyen:2009fk,Hanahan:2011ly}. The process of invasion and the modes of motility associated with it are a subject of intense study. Depending on cancer type, cells may break away from the primary tumour and migrate to the blood vessels as single cells, collectively as detached clusters or as multicellular, three-dimensional invasive strands~\cite{Friedl:2009fk,Friedl:2012fk}. Undoubtedly, the acquisition of mesenchymal traits by the invasive cancer cells, together with cell motility and the expression of proteases, play central roles in many of these cases and particularly in malignant tumours~\cite{Thiery:2002uq,Friedl:2008uq,Thiery:2009kx,Kessenbrock:2010ly}. However, undulations and protrusions of the epithelium into the stroma are commonly found in benign tumours and even in healthy epithelia, where the basement membrane separating the two tissues remains intact. As the tissue progresses to higher grades of malignancy, the number of dividing layers within the epithelium increases and such protrusions typically grow in size~\cite{tavassoli2003pag,weinberg2007bc}. This is the case for example in cervical intraepithelial neoplasia~\cite{tavassoli2003pag,park1998coexistence} and in the epithelial dysplasia of the oral mucosa~\cite{bouquot2006epithelial,Tsai:2008ys}. In this work, we investigate a potential mechanism underlying these commonly observed, yet striking morphological features of stratified epithelia and carcinomas, which relies on proliferation-induced mechanical stresses without invoking proper cell motility. The mechanics of this process is of particular interest, considering that these protrusions arise in epithelia whose apical surfaces are free of stress, such that it is non-trivial how an increased proliferation rate alone like that observed in neoplastic tissues can lead to invasive protrusions into the stroma.

In a previous work~\cite{Basan:2011fk}, we have demonstrated the existence of a mechanical instability based entirely on the flow caused by cell renewal in the epithelium in combination with its viscosity due to cell-cell adhesion. We showed that when the epithelium-stroma interface is displaced from an originally flat configuration, the excess of cell division above a nascent protrusion creates a differential flow of cells, and the resulting viscous shear stress drives the protrusion further. A finite-wavelength instability develops from the combination of surface-stabilizing factors such as interfacial tension and mechanical resistance of the stroma on the one hand, and destabilizing factors enhancing cell-division driven flows and viscous shear stresses on the other hand. In our previous work, we studied the two limits of a purely elastic or purely viscous rheology of the stroma. The present paper serves both as a more detailed presentation of this instability as well as to propose two important extensions of our earlier work. First, we study the instability in the general case of a viscoelastic stroma and investigate the transition between the two regimes presented previously, which we recover as limits. Second, and most importantly, we include a dependence of the cell-division rate in the epithelium on the local concentration of a substance necessary for cell division---representing nutrients, growth factors  or oxygen---which diffuse through the epithelium from the stroma where it is delivered by blood vessels~\cite{young2006wheater}. Compared with our previous study where the rate of cell renewal in the epithelium was pre-defined as a function of distance from the stroma, this coupling leads to a significant enhancement of the instability by a diffusion-limitation mechanism, which is similar to mechanisms leading to instabilities in other contexts~\cite{mullins1964stability,langer1980instabilities,Langer:1989fk,ben1990formation}. We comment further in the discussion on the comparison with classical instabilities known from non-equilibrium physics  as well as on other dynamical instabilities observed in living systems such as bacterial colonies.

In tissues, mechanical instabilities may play a role in the morphogenesis of certain shapes and patterns exhibited by growing cell populations. For example, differential growth has been proposed as a mechanism underlying the large-scale looping morphology of the gut~\cite{Savin:2011cr}, and a buckling instability of a monolayered epithelium has been suggested for the formation of vili and crypts in the colon~\cite{Drasdo:2000pi,Hannezo:2011qf}. In the case of multilayered epithelia, a buckling instability of the basal layer of the fetal epidermis has been proposed to be at the origin of the formation of apidermal ridges~\cite{Kucken:2005vn}. The complex network of finger-like protrusions at the dermal-epidermal junction of human skin has been proposed to result from incompatible growth of elastic tissue layers~\cite{ciarletta2011papillary}. Similar instabilities have been investigated in different geometries, such as that of a growing elastic shell with differential growth~\cite{ben2005growth,Dervaux:2008ff,Muller:2008qa,Liang:2009fk,Liang:2011uq} or that of a tube~\cite{hannezo2012mechanical}. These works are part of the broader field related to the role of mechanics for growing tissues and of their microenvironment~\cite{Nelson:2005vn,Hufnagel:2007ly,Mammoto:2010bh,jones2012modeling}.

In the specific case of tumour growth, pattern formation has been reported and studied theoretically using reaction-diffusion descriptions of the supply of nutrients, oxygen or growth factors~\cite{tracqui2009biophysical}. Such a coupling affects tumour-growth dynamics~\cite{Ciarletta:2013kx}, as well as tumour shape and interfacial structure~\cite{Ferreira:2002il}. Fingering instabilities may develop depending on the degree and spatial structure of vascularization~\cite{Cristini:2003zt}. Taking into account the mechanical properties of the growing tissue and that of its microenvironment, residual stresses~\cite{ambrosi2002mechanics} and fingering instabilities may or may not develop~\cite{Macklin:2007cr}. In particular, weakened cell-cell adhesions and cell-matrix adhesion have been proposed to favor such instabilities~\cite{byrne1996modelling,Frieboes:2006gb}. Instabilities may also result from a combination of pushing forces due to cell proliferation and pulling forces of the tissue on either a substrate or the extra-cellular matrix in the case of a three-dimensional, multi-cellular spheroid~\cite{drasdo2012modeling}. However, the coupling of growth dynamics to the diffusion field of nutrients has never been investigated within the framework of the instability proposed here, which differs from the aforementioned studies because it is based on a purely hydrodynamic description of the tissue and because it arises in the specific geometry of an epithelium with a free apical surface rather than that of a tumor spheroid or a two-dimensional tissue spreading on a substrate. Because this additional diffusion-dependent mechanism can lead to an unstable epithelium-stroma interface over a broader range of parameters than the instability reported previously~\cite{Basan:2011fk}, we expect this mechanism to potentially play an important role in the generation of undulated patterns as they occur in real epithelial tissues.

In this paper, we present six different versions of a linear stability analysis of the epithelium-stroma interface for different rheologies of the stroma and cell-renewal functions in the epithelium. In section~2, we start by presenting the common properties of the six models that will be studied in the paper. The associated detailed equations can be found in the Supplementary Information. Models~1 and 2, where the cell-production function in the epithelium is pre-defined and the stroma is either purely elastic (model~1) or purely viscous (model~2) have already been presented in~\cite{Basan:2011fk}. The results associated with these models are therefore presented in the Supplementary Information including some novel aspects as compared with our previous work~\cite{Basan:2011fk}. In section~3, we generalize the two previous cases to that of a viscoelastic rheology of the stroma (model~3), and compare the mode structures associated with these three different models. Section~4 is devoted to the description of the coupling of cell division in the epithelium to the local concentration of available nutrients, which diffuse from the stroma and are consumed by the epithelial cells. In sections~5, 6 and 7, we present the resulting stability analysis successively in the case of an elastic (model~4), a viscous (model~5) and a viscoelastic (model~6) stroma. Domains of validity, similarities, and differences between the various models are discussed throughout the presentation. We finally discuss the similarites and differences between the results presented in this study and other instabilities know from the literature.

\section{Description of models~1-3}

We start this section by presenting the geometry of the system, which underlies the six different models presented in the paper. We then present the structure of models~1-3, where the cell-production function in the epithelium is pre-imposed as a function of distance from the epithelium-stroma interface. The structure of the models is just described here, and the full set of equations can be found in the Supplementary Information. The results associated with models 1 and 2, already present for the main part in our previous report~\cite{Basan:2011fk}, can be found in the Supplementary Information. We present here in more details the results obtained with model 3, which constitutes a first extension of our previous work.

\subsection{Description of the geometry}\label{subsecModelGeom}

\begin{figure}[ht]
\scalebox{0.3}{
\includegraphics{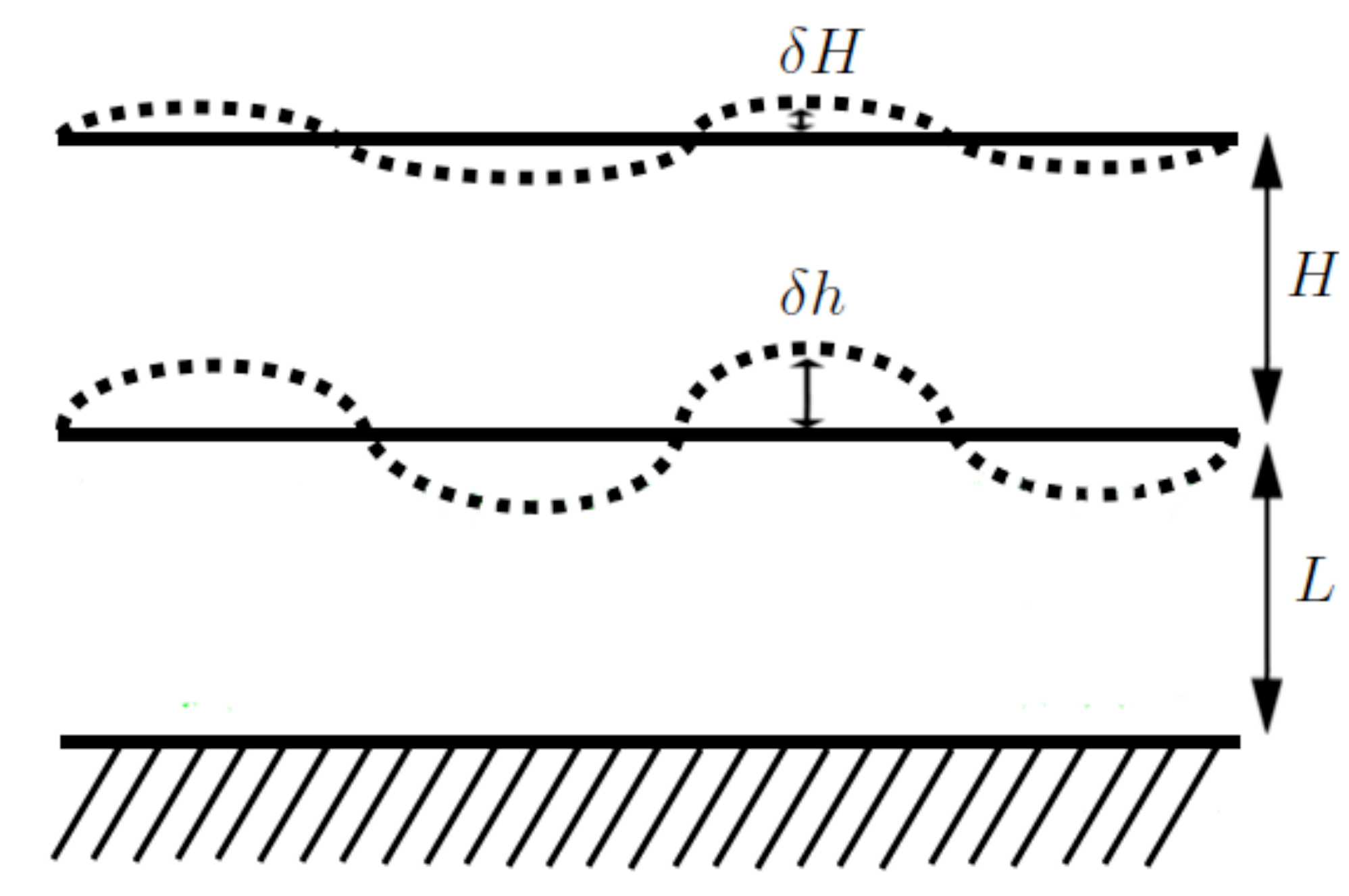}
}
\caption
{\label{Fig_Geometry}
Schematic representation of the geometry considered in this study. An epithelium of thickness $H$ is adjacent to a stroma of thickness $L$, itself anchored to a static, rigid structure at its opposite side. Perturbations of the epithelium-stroma interface and of the apical surface of the epithelium have infinitesimally small amplitudes, labelled $\delta h$ and $\delta H$, respectively.}
\end{figure}

We consider an epithelium of thickness $H$ adjacent to a stroma of thickness $L$, infinitely extended in directions $x$ and $y$. In the $z$-axis, we assume a rigid wall at $z=0$ to which the stroma is anchored, followed by the stroma, then the epithelium starting at $z=L$, and finally the apical surface of the epithelium directly adjacent to a lumen at $z=L+H$ (figure~\ref{Fig_Geometry}).
These coordinates correspond to the flat configuration and are strictly valid only in the unperturbed, steady state. We study the stability of these interfaces under a given arbitrary but infinitesimally small perturbation. In this case, the system of equations can be linearized, and it is sufficient to study a deformation $\delta h$ of the epithelium-stroma interface translationally invariant in the $y$ direction with the complex form $\delta h(x,t) = \delta h_0 \exp(\omega t + {\rm i} q x)$. Similarly, a perturbation of the apical surface of the epithelium can be written $\delta H(x,t) = \delta H_0 \exp(\omega t + {\rm i} q x)$. The stability of the system is then given by the dispersion relations $\omega(q)$ for each of the relaxation modes.

\subsection{Description of the models}\label{descModel}

In general, for length and time scales large compared with cell size and characteristic times of individual cellular processes, biological tissues can be described as continuous media with a potentially complex rheology, intermediate between those of liquids and solids~\cite{Forgacs:1998bh,verdier2003rheological,Marmottant:2009lh,Gonzalez-Rodriguez:2012fk}. In our previous work~\cite{Basan:2011fk}, as in other theoretical studies of tissue growth~\cite{bittig2008dynamics,Behrndt:2012fk}, a viscous rheology was chosen to describe the epithelium, based on the assumption that cell-cell junction rearrangements lead to the relaxation of static stresses on long timescales~\cite{Gordon:1972bh,Frisch:1986xb} and to effective cell flows~\cite{Eaton:2011fk}. This choice is motivated by the observation that relaxation times are of the order of tens of minutes to several hours for embryonic tissues in compression-plate experiments~\cite{Forgacs:1998bh,Marmottant:2009lh} as well as in pipette-aspiration experiments~\cite{Guevorkian:2010ye}. Such results have also been proven accurate in the specific case of carcinomas, which show an almost complete stress relaxation under imposed sequential strains~\cite{Netti:2000qo}. In addition, this choice is supported by the experimentally observed presence of surface tension at tissue boundaries~\cite{Foty:1996cr,Mgharbel:2009qf,Guevorkian:2010ye,Gonzalez-Rodriguez:2012fk}, which may drive cell sorting and certain morphogenetic movements~\cite{Foty:1996cr,Lecuit:2007fu,Krieg:2008qf}. On timescales much longer than the characteristic time of cell turnover in the tissue, repeated cycles of cell-division and apoptosis can lead to the same result even for tissues that behave like elastic media or yield-stress fluids on short timescales~\cite{Ranft:2010uq,Basan:2011fl}. We therefore model the epithelium as a viscous fluid of shear viscosity $\eta$, which we take incompressible for the sake of simplicity, but with a rate of material production $k_{\rm p}$ corresponding to cellular duplication minus cellular death. In models 1-3, this rate is pre-imposed and takes the form of a single exponential function with a characteristic decay length $l$, whose amplitude is determined by a rate parameter $k$ (detailed equations can be found in the Supplementary Information, section~1).   

In contrast to the epithelium where cells are confluent, the stroma is made of a network of proteoglycans, collagen and elastin fibers, within which cells---mostly fibroblasts--- are sparse~\cite{cellB, young2006wheater}. Fibroblasts constantly remodel the stromal fibers, which can therefore elastically resist deformation only on short and intermediate timescales but eventually reorganize and follow imposed deformations on long timescales. In addition, matrix metalloproteinases expressed either by the advancing tumour cells or directly by tumour-associated stromal cells can enhance this process by digesting the filaments of the stroma~\cite{weinberg2007bc,Friedl:2008uq,Kessenbrock:2010ly,Hanahan:2011ly}. The stroma should therefore be thought of more as a viscoelastic material than a purely elastic or viscous medium, with a characteristic timescale above which it reorganizes and flows. Here, three different descriptions are envisaged, namely those of an elastic solid with shear modulus $\mu$ (model 1), a Newtonian viscous fluid with shear viscosity $\eta_{\rm s}$ (model 2) or a viscoelastic material described by a Maxwell model with a single relaxation time $\tau=\eta_{\rm s}/\mu$ (model 3). In each case, the medium is supposed incompressible for the sake of simplicity.

Boundary conditions are as follows: At the apical surface of the epithelium, the total stress in contact with the lumen vanishes and the cell velocity is equal to the time derivative of the apical-surface location. Taking into account the epithelium apical surface tension~$\gamma_{\rm a}$, the normal component of the stress is given by the Laplace pressure, and its tangential component vanishes. At the epithelium-stroma interface, the discontinuity of the normal component of the stress is given by Laplace's law with interfacial tension $\gamma_{\rm i}$, and the tangential components of the stress are continuous and equal to a finite surface-friction term with coefficient $\xi$. The normal component of the velocity is continuous and the stroma displacement in the $z$-direction is equal to $\delta h$. At the bottom of the stroma, the displacement vanishes. All the corresponding equations are detailed in the Supplementary Information, section~1.

\section{Results for model 3}

We present here the first generalization of our previous study~\cite{Basan:2011fk}, namely that of a viscoelastic stroma with the cell-production rate function in the epithelium pre-defined as a function of distance from the epithelium-stroma interface. This model~3 encompasses both of the previously exposed models 1 and 2 as limit cases, where the stroma was either purely elastic or purely viscous~\cite{Basan:2011fk} (see also the Supplementary Information, sections~2 and 3). We shall see the consequences of this more general model and how we recover the two cases studied before in particular regimes.

\subsection{Mode structure and comparison with models~1 and 2}

We obtain four different relaxation modes. This can be understood from the structure of the equations describing the boundary conditions, where time derivatives appear four times: once in the continuity condition for the velocity at the apical surface of the epithelium, and three times in the boundary conditions at the epithelium-stroma interface (Supplementary Information, equations~(14) and (17)). We get four relaxation modes rather than three as in the case of an elastic stroma because of the additional relaxation timescale of the viscoelastic rheology here.

\begin{figure}[ht]
\scalebox{0.86}{
\includegraphics{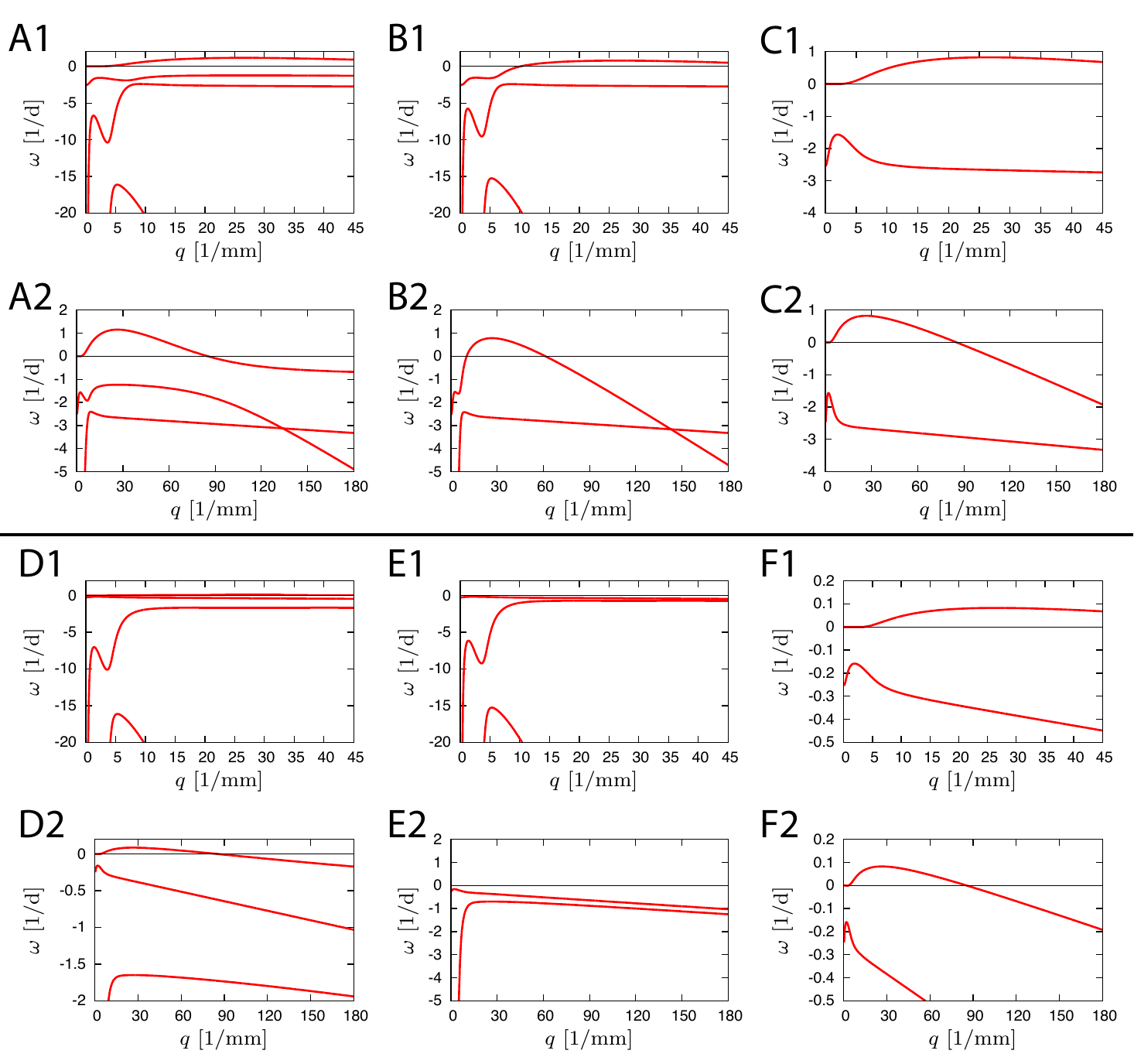}
}
\caption
{\label{Fig_ViscEl_Bis}
Relaxation modes $\omega$ as a function of the wave number $q$ for model~3, and comparison with the results given respectively by models 1 and 2. (A1) The relaxation modes corresponding to the viscoelastic model~3 are plotted for parameter values that are identical to those of figure 1A of the Supplementary Information for model~1, to which must be added the stroma viscosity $\eta_{\rm s}=10$~MPa$\cdot$s. Combined with the elastic modulus of the stroma $\mu=100$ Pa, this gives an inverse viscoelastic relaxation time $\tau^{-1}$ of 0.86 per day. (A2) The span of wavenumber values has been augmented in order to visualize the convergence of the most unstable mode to this inverse relaxation time, and the range of values displayed for $\omega$ has been shortened to zoom into the relevant region ; consequently, only three of the four modes appear here. (B) and (C) The corresponding plots are shown respectively for the elastic and viscous models~1 and 2, with the same parameters and the same panel structure. (D1) The relaxation modes corresponding to the viscoelastic model~3 are plotted for a set of parameters that is now identical to that used in figure 2A of the Supplementary Information for model~2, except for a different stroma viscosity and the addition of the elastic modulus $\mu=100$~Pa. The stroma viscosity takes the value $\eta_{\rm s}=10$~MPa$\cdot$s, chosen such that the inverse viscoelastic relaxation time $\tau^{-1}$ is maintained at 0.86 per day. (D2) The span of wave number values has been augmented together with a zoom into the region of small relaxation rates to visualize more clearly the two upper modes. (E) and (F) The corresponding plots are shown for the respective cases of models~1 and 2, with the same parameters and the same panel structure.}
\end{figure}

We present in figure~\ref{Fig_ViscEl_Bis} the structure of the relaxation modes as a function of the wave number $q$, and compare the present results to those of the purely elastic and purely viscous models~1 and 2, using the same set of parameters.
We choose the two tissue viscosities to be equal to 10~MPa$\cdot$s for models~2 and 3, and fix the viscoelastic relaxation rate $\tau^{-1}$ to 0.86 per day for model~3. This corresponds to an elastic modulus $\mu$ of 100~Pa for models~1 and 3. The structure of the figure is as follows: Plots corresponding to a first set of parameters are shown in figures~\ref{Fig_ViscEl_Bis}A to \ref{Fig_ViscEl_Bis}C, and for a second set in figures~\ref{Fig_ViscEl_Bis}D to \ref{Fig_ViscEl_Bis}F. For each of these parameter sets, the first column corresponds to the viscoelastic model~3, the second column to the elastic model~1, and the third one to the viscous model 2. In addition, for each of the mode structures computed, the corresponding plots are displayed in a range of wavevector values spanning 0 to 45 per millimeter in the upper subpanels (indices 1), and 0 to 180 per millimeter in the lower subpanels (indices 2). The range of omega values in the vertical axes is adapted to each individual case to show different relevant parts of the mode structure.

In the first set of graphs, we can see from the curves associated with model~3 in figure~\ref{Fig_ViscEl_Bis}A that one of the modes relaxes toward the inverse viscoelastic relaxation time $\tau^{-1}$ of 0.86 per day. In figure~\ref{Fig_ViscEl_Bis}B, the associated plots for the elastic model~1 almost identically reproduce the two lower modes, which are associated with timescales shorter than $\tau$ and therefore correspond to the elastic limit. We also recognize features of the two upper modes of the full viscoelastic model in both plots associated with models~1 and 2 (figures~\ref{Fig_ViscEl_Bis}B and \ref{Fig_ViscEl_Bis}C, respectively), although here these features are shared between the two different models depending on the range of wave numbers. In the second set of graphs (figures~\ref{Fig_ViscEl_Bis}D to \ref{Fig_ViscEl_Bis}F), we can appreciate the similarities between the viscoelastic and elastic relaxation modes whenever these are associated with short timescales, and between the viscoelastic and vicous relaxation modes in the other limit.

\subsection{Asymptotic behaviours of the modes}

It is instructive to look at the analytic expansions of the different modes in the respective limits of large and small wave numbers $q$. In the limit of large wave numbers, the modes associated with the epithelium-stroma interface decouple from those associated with the apical surface of the epithelium, since their characteristic decay lengths are of order $q^{-1}$, which is much smaller than $H$ in this limit. Their asymptotic expressions up to constant order in a development in powers of $q^{-1}$ read:
\newpage
\begin{eqnarray}\label{largeqModesViscoElSimple}
\omega_1 &\simeq& -\frac{\gamma_{\rm i}}{2\eta}q-\frac{\mu}{\eta}+k-k_0\nonumber\\
\omega_2 &\simeq& -\frac{\gamma_{\rm a}}{2\eta}q+k{\rm e}^{-H/l}-k_0\nonumber\\
\omega_3 &\simeq& -2 \frac{\mu}{\xi}q-\frac{(\eta+ \eta_{\rm s})\mu}{\eta\eta_{\rm s}}\nonumber\\
\omega_4 &\simeq& 0-\frac{\mu}{\eta_{\rm s}}.
\end{eqnarray}
Among these, the first three expressions correspond to short characteristic times. We therefore expect their behaviours to be similar to those obtained in the case of an elastic stroma of modulus $\mu$ as already discussed in~\cite{Basan:2011fk} and summarized in the Supplementary Information (see equation~(18)). This is indeed the case for all of these three modes to leading order, and for the two first ones even up to constant order. In the third mode, the term of constant order mixes the epithelium and stroma viscosities, and therefore departs from the simpler $\mu/\eta$ term present in the elastic case. The additional fourth mode converges toward the inverse relaxation time $\tau^{-1}=\mu/\eta_{\rm s}$ of the viscoelastic stroma, as illustrated in figure~\ref{Fig_ViscEl_Bis}A2.

In the limit of small wave numbers, systematic expansions to leading order read:
\begin{eqnarray}\label{smallqModesViscoElSimple}
\omega_1 &\simeq& -\frac{36\mu}{\eta}\frac{1}{H^3L^3}\frac{1}{q^6}\nonumber\\
\omega_2 &\simeq& -\frac{\mu}{4\eta}\frac{1}{HL}\frac{1}{q^2}\nonumber\\
\omega_3 &\simeq& k{\rm e}^{-H/l}-k_0\nonumber\\
\omega_4 &\simeq& -\frac{L^2\left[3H\gamma_{\rm a}+2L(\gamma_{\rm a}+\gamma_{\rm i})\right]}{6\eta_{\rm s}}q^4.
\end{eqnarray}
Here, the expressions mix contributions coming from the entire system. The asymptotic expressions of the two first modes are identical to those already obtained in the case of an elastic stroma, and the one of the fourth mode corresponds to that obtained in the case of a viscous stroma (equations~(19) and (21), Supplementary Information). This can be easily understood from the observation that the two first limits correspond to short-time dynamics and the fourth one to long-time dynamics. As for the third asymptotic expression, it is common to the three models. This is because this limit comes from pure mass-conservation in the epithelium and is therefore independent of the stroma rheology. One can indeed recover this limit by integrating the continuity equation at $q=0$ over the height of the epithelium to leading order in the perturbations. Expressions to next-to-leading order can be found in the Supplementary Information, section~5.

\newpage

\section{Nutrient-diffusion dynamics}

As discussed in the introduction, the first three models exposed above and in the Supplementary Information sections~1-3 rely on a specific pre-defined function for the rate of cell division minus apoptosis in the epithelium, characterized by an exponential decay with a single characteristic length $l$ (equations~(2) and (3), Supplementary Information). This description is motivated by the fact that the supply of nutrients and growth factors in the epithelium comes from diffusion from the stroma. However, the function used above was set {\it a priori}, and to be more accurate, nutrient diffusion needs to be solved explicitly. Here, we solve the entire system of equations, where nutrients are produced at a given location in the stroma, diffuse in both tissues with tissue-specific diffusion constants, and can leak from the epithelium at its apical surface if they are in excess. They are consumed by the epithelial cells and influence their division rate.

\subsection{Bulk equations}

The equations describing the epithelium are still given by equation~(1) of the Supplementary Information, but we now couple the cell-production rate $k_{\rm p}$ to the nutrient density $\rho$. In the absence of further detailed knowledge about this coupling, we assume a linear dependence of $k_{\rm p}$ on $\rho$:
\begin{equation}\label{CellProdFctNutrients}
k_{\rm p} = \kappa_1\rho-\kappa_0,
\end{equation}
where $\kappa_1$ and $\kappa_0$ are effective phenomenological coefficients. These coefficients take positive values, since we expect the cell-division rate to increase as a function of $\rho$, as well as the overall cell population to starve and progressively die in the absence of nutrients. The nutrient-density function $\rho$ is determined by the following diffusion-consumption equations. In the epithelium, nutrients diffuse with a coefficent $D$ and are consumed by the cells with a rate $c$:
\begin{equation}\label{NutDynEpi}
\partial_t\rho=D\Delta\rho-c\rho.
\end{equation}
In the stroma, nutrients have the density $\rho_{\rm s}$ and diffuse with a coefficent $D_{\rm s}$, without being consumed:
\begin{equation}\label{NutDynStroma}
\partial_t\rho_{\rm s}=D_{\rm s}\Delta\rho_{\rm s}.
\end{equation}
For the stroma, we again consider three different versions of the rheology as investigated above: an elastic rheology for model~4, a viscous one for model~5 and a viscoelastic one for model~6. The associated equations are presented in the Supplementary Information, section~1.

\subsection{Boundary conditions}\label{nutrientBC}

Mechanical boundary conditions are identical to those presented in the previous sections, but we need to specify the boundary conditions for the nutrient fields $\rho$ and $\rho_{\rm s}$. We assume a fixed concentration of nutrients $ \bar{\rho}_0$ a distance $d$ away from the epithelium-stroma interface in the stroma compartment:
\begin{equation}
{\rho_{\rm s}}_{|z=L-d}= \bar{\rho}_0.
\end{equation}
At the epithelium-stroma interface, the density as well as the flux of nutrients in the direction perpendicular to the local interface are continuous:
\begin{eqnarray}\label{NutInterfBC}
\rho_{|\rm interface} &=& {\rho_{\rm s}}_{|\rm interface}\nonumber\\
D\partial_\perp\rho_{|\rm interface} &=& D_{\rm s} \partial_\perp{\rho_{\rm s}}_{|\rm interface}.
\end{eqnarray}
Here `$\partial_\perp$' stands for the partial derivative in the direction perpendicular to the local interface, oriented positively from the stroma toward the epithelium. At the apical surface of the epithelium, we allow for a potential leakage of nutrients from the epithelium into the lumen. Since we expect this leakage to increase with the local nutrient concentration, we write to linear order that the nutrient flux is proportional to this concentration locally:
\begin{equation}
-D\partial_\perp\rho_{|\rm apical} = v_{\rm off}\rho_{|\rm apical}.
\end{equation}

\section{Results for an elastic stroma: model 4}\label{model4}

We present here the mode structure for model 4, where the stroma is treated as an elastic, incompressible material. The expressions of the steady-state solutions for the nutrient-concentration, velocity and stress fields in the epithelium as well as the perturbed equations to linear order can be found in the Supplementary Information, section~4. To solve this system of equations, we consider the case where nutrient diffusion is much faster than the characteristic relaxation of the system, both in the stroma ($\omega\ll D_{\rm s}q^2$) as well as in the epithelium ($\omega\ll Dq^2+c$). This approximation is valid sufficiently close to the instability threshold and is a standard approximation of the treatment of diffusion-limited interface dynamics~\cite{mullins1964stability,langer1980instabilities}. In this regime, we can solve the equations describing nutrient diffusion at steady state first, and then substitute the obtained solutions into the mechanical equations.

\subsection{Mode structure and influence of diffusion}\label{subse_Mode_Nutrients_El}

\begin{figure}[ht]
\scalebox{0.95}{
\includegraphics{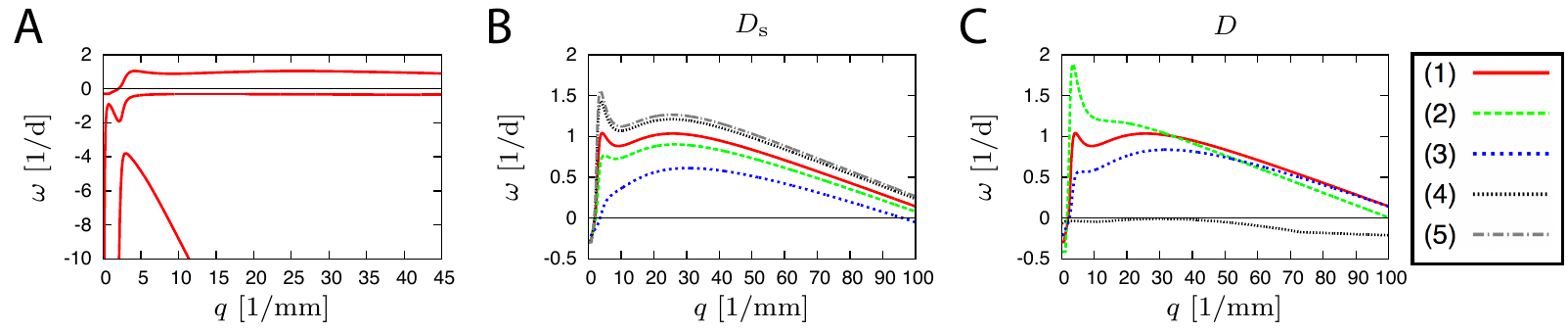}
}
\caption
{\label{Fig_Elastic_Nutrients}
Relaxation modes for model~4 as a function of nutrient diffusion. (A) Parameters are as follows: $\eta=30$~MPa$\cdot$s, $\mu=50$~Pa, $\gamma_{\rm i}=10$~mN$\cdot$m$^{-1}$, $\gamma_{\rm a}=1$~mN$\cdot$m$^{-1}$, $\xi=10$~GPa$\cdot$s$\cdot$m$^{-1}$, $H=300$~$\mu$m, $L=600$~$\mu$m, $d=75$~$\mu$m, $\kappa_1=100$~$10^{-6}$~m$^3\cdot$s$^{-1}$, $D=200$~$10^{-12}$~m$^2\cdot$s$^{-1}$, $D_{\rm s}=100$~$10^{-12}$~m$^2\cdot$s$^{-1}$, $c=0.1$~s$^{-1}$, $\bar{\rho}_0=1$~m$^{-3}$ and $v_{\rm off}=0$~m$\cdot$s$^{-1}$. All three relaxation modes are shown, of which only one goes unstable and presents a clear maximum at low-$q$ values.
(B) The most unstable mode is shown for the same parameter set as in panel (A) for the default curve (1), and $D_{\rm s}$ is varied in the other curves. It takes the following values: (1) $100$~$10^{-12}$~m$^2\cdot$s$^{-1}$, (2) $200$~$10^{-12}$~m$^2\cdot$s$^{-1}$, (3) $1000$~$10^{-12}$~m$^2\cdot$s$^{-1}$, (4) $20$~$10^{-12}$~m$^2\cdot$s$^{-1}$, (5) $2$~$10^{-12}$~m$^2\cdot$s$^{-1}$. (C) A similar analysis is presented while varying $D$, which takes the following values: (1) $200$~$10^{-12}$~m$^2\cdot$s$^{-1}$, (2) $500$~$10^{-12}$~m$^2\cdot$s$^{-1}$, (3) $100$~$10^{-12}$~m$^2\cdot$s$^{-1}$, (4) $10$~$10^{-12}$~m$^2\cdot$s$^{-1}$. In both panels, $\bar{\rho}_0$ is changed as $D_{\rm s}$ or $D$ are varied to keep the concentration of nutrients unchanged at the epithelium-stroma interface. Plots are coded both in colour as well as linestyles as indicated on the right of the figure.}
\end{figure}

The number of modes that we get is identical to the one obtained with a pre-imposed function for the production of cells, since the number of modes is prescribed by the structure of the mechanical boundary conditions. In the particular case of an elastic stroma, we get three relaxation modes, which can be seen from the expressions of the boundary conditions, where time derivatives appear three times (equations~(14) and (15), Supplementary Information). In figure~\ref{Fig_Elastic_Nutrients}, we study the influence of nutrient diffusion on the structure of these modes.
In figure~\ref{Fig_Elastic_Nutrients}A, we show all three relaxation modes. The main qualitative difference between the present curves and those associated with model~1 (figure~1, Supplementary Information) is the presence of a clear maximum at low values of the wave number $q$ (here around 5 mm$^{-1}$), that is at long wavelengths, in addition to the other broader maximum at smaller wavelength already present in the previous model (here around $q$=30 mm$^{-1}$). To test whether this new maximum is indeed controlled by nutrient diffusion, we vary in figures~\ref{Fig_Elastic_Nutrients}B and \ref{Fig_Elastic_Nutrients}C the values of the diffusion coefficients of nutrients respectively in the stroma and in the epithelium. In doing so, we change the value of the nutrient concentration at its production location ($\bar{\rho}_0$) in order to keep a constant concentration at the epithelium-stroma interface in the unperturbed steady state. This allows us to decouple the influence of diffusion from that of the overall amount of nutrient supply. We see in figure~\ref{Fig_Elastic_Nutrients}B that increasing nutrient diffusion in the stroma makes the maximum at low $q$ disappear, and that decreasing it instead makes the curve saturate toward a maximum limit curve with a pronounced peak. This behaviour is in qualitative agreement with that of an instability controlled by diffusion limitation~\cite{mullins1964stability,langer1980instabilities,ben1990formation}. The situation is different in terms of the variation of $D$, the diffusion coefficient in the epithelium, as illustrated in figure~\ref{Fig_Elastic_Nutrients}C. In this case indeed, increasing $D$ favors the instability and decreasing it makes the system stable. This stems from the fact that slowing down diffusion in the epithelium makes the layer of dividing cells very thin. This in turns lowers the driving force for the instability, which originates from differential cell flows in the epithelium due to inhomogeneous cell divisions.

\begin{figure}[ht]
\scalebox{1.5}{
\includegraphics{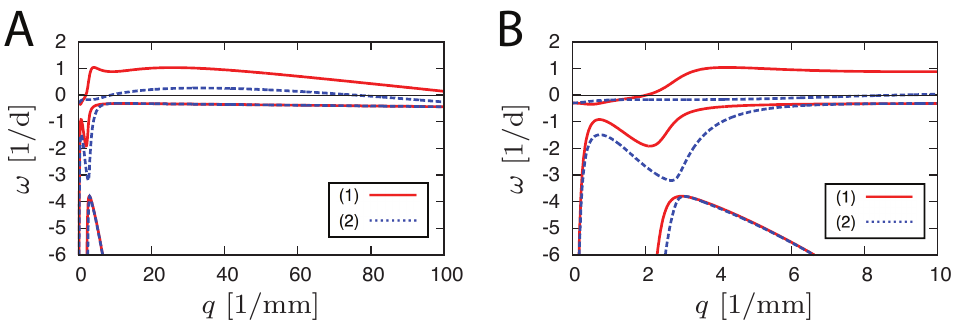}
}
\caption
{\label{Fig_Elastic_Nutrients_Comparison}
Comparison of the modes obtained with model~4 (linestyle (1)) and with model~1 (linestyle (2)), following the procedure described in the text. The parameters are those of figure~\ref{Fig_Elastic_Nutrients}A, together with the fitting parameters $k=2$ division per day and $l$=44 $\mu$m for model~1. Panels (A) and (B) present the same modes but over two different wavevector ranges.}
\end{figure}

We now compare the results of the two models more directly, namely those obtained with a pre-defined cell-production function (model~1) {\it versus} those obtained while coupling the cell-production function to nutrient diffusion (model~4). In order to do so, we derive from model~4 a cell-production rate as a function of $z$ at steady state. We then fit the cell-production rate of model~1 to that function to generate the associated relaxation modes. As illustrated in figure~4 of the Supplementary Information, the fitting procedure gives a nearly perfect agreement between the two rate functions. We can therefore assume that the cell-production profiles are identical in the unperturbed situation for both models, and that all the differences that are observed are due to the effect of being coupled to nutrient diffusion to first order in the case of model~4 {\it versus} keeping the cell production unchanged in the case of model~1. We can see in figure~\ref{Fig_Elastic_Nutrients_Comparison} that coupling to nutrient diffusion overall increases the instability and introduces an additional maximum for the most unstable mode at a low $q$ as commented above.
The remaining of the mode structure is unchanged and the stable modes are nearly identical in both models. In particular, in both the large- and low-$q$ limits, we have the same asymptotic behaviours for the three modes with both models, as we now discuss in paragraph~\ref{subse_ModeAs_Nutrients_El}.

\subsection{Asymptotic behaviours of the modes}\label{subse_ModeAs_Nutrients_El}

To leading order, the large-$q$ expansions of the three modes are identical to those obtained for model~1 and given by equation~(18) of the Supplementary Information. This is also the case in the small-$q$ limit for the two diverging modes of model 1, given by equation~(19) of the Supplementary Information. This stems from the fact that these regimes correspond to a fast dynamics, where pure mechanics is at play without feeling the influence of the relatively slow cell-division events. However, the constant-order terms of these expansions do depend on the particular form of the cell-production function. The exact expressions have been derived but are very long and do not yield any particular physical insight in their full extent. They are therefore not presented in this paper. In figure~\ref{Fig_Elastic_Nutrients_Comparison} of the previous paragraph, the fitting procedure insures that the limit of the upper mode is the same for both versions of the model.

\section{Discussion for a viscous stroma: model~5}

We now investigate the equivalent model for a viscous stroma.

\subsection{Effect of nutrient diffusion on the mode structure}

\begin{figure}[ht]
\scalebox{0.95}{
\includegraphics{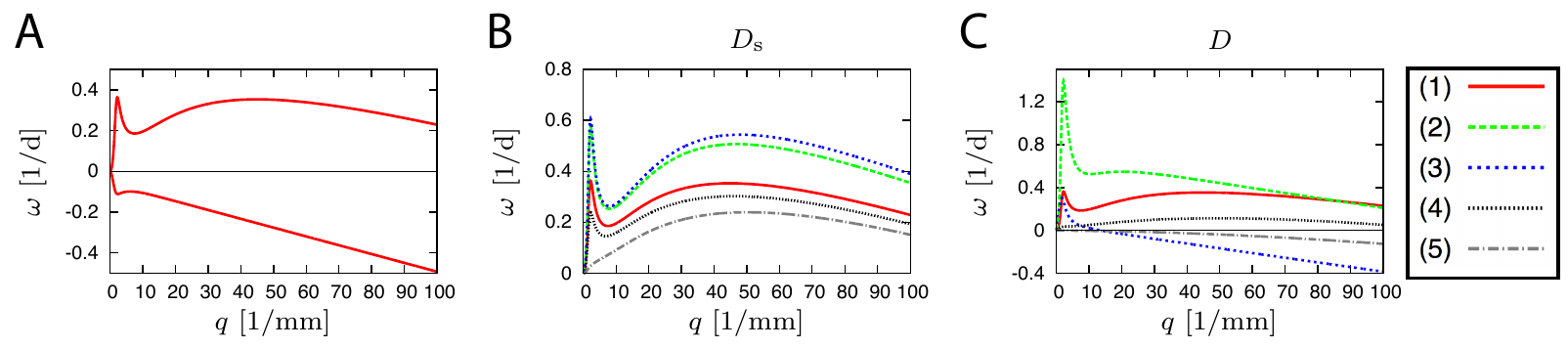}
}
\caption
{\label{Fig_Viscous_Nutrients}
Relaxation modes for model~5 as a function of nutrient diffusion. (A) Parameters are as follows: $\eta=10$~MPa$\cdot$s, $\eta_{\rm s}=10$~kPa$\cdot$s, $\gamma_{\rm i}=1$~mN$\cdot$m$^{-1}$, $\gamma_{\rm a}=1$~mN$\cdot$m$^{-1}$, $\xi=10$~GPa$\cdot$s$\cdot$m$^{-1}$, $H=L=300$~$\mu$m, $d=100$~$\mu$m, $\kappa_1=100$~$10^{-6}$~m$^3\cdot$s$^{-1}$, $D=100$~$10^{-12}$~m$^2\cdot$s$^{-1}$, $D_{\rm s}=200$~$10^{-12}$~m$^2\cdot$s$^{-1}$, $c=0.2$~s$^{-1}$, $\bar{\rho}_0=0.3$~m$^{-3}$ and $v_{\rm off}=0$~m$\cdot$s$^{-1}$. Both relaxation modes are shown, of which only one goes unstable and presents a clear maximum at low-$q$ values.
(B) The most unstable mode is shown for the same parameter set as in panel (A) for curve (1), and $D_{\rm s}$ is varied in the other curves. As for figure~\ref{Fig_Elastic_Nutrients}, the value of $\bar{\rho}_0$ is set accordingly to keep the concentration of nutrients unchanged at the epithelium-stroma interface. Values of $D_{\rm s}$ are: (1) $200$~$10^{-12}$~m$^2\cdot$s$^{-1}$, (2) $20$~$10^{-12}$~m$^2\cdot$s$^{-1}$, (3) $2$~$10^{-12}$~m$^2\cdot$s$^{-1}$, (4) $400$~$10^{-12}$~m$^2\cdot$s$^{-1}$, (5) $2000$~$10^{-12}$~m$^2\cdot$s$^{-1}$. (C) The same procedure is applied to study the variation of $D$, which takes the values: (1) $100$~$10^{-12}$~m$^2\cdot$s$^{-1}$, (2) $1000$~$10^{-12}$~m$^2\cdot$s$^{-1}$, (3) $100$~$10^{-9}$~m$^2\cdot$s$^{-1}$, (4) $20$~$10^{-12}$~m$^2\cdot$s$^{-1}$, (5) $5$~$10^{-12}$~m$^2\cdot$s$^{-1}$.}
\end{figure}

As for model~4, we study in figure~\ref{Fig_Viscous_Nutrients} the influence of the dynamics of nutrient diffusion.
Here, we have two relaxation modes, since as compared with the elastic case, we loose the mode that originates from the tangential stress-balance condition at the epithelium-stroma interface (see equation~(16), Supplementary Information). Similarly to the previous case, the main difference with model~2 is the presence of an extra maximum of the unstable mode at long wavelengths (figure~\ref{Fig_Viscous_Nutrients}A here and figure~2A of the Supplementary Information). In figure~\ref{Fig_Viscous_Nutrients}B, we study the behaviour of this maximum as we vary the nutrient diffusion constant in the stroma $D_{\rm s}$ while keeping the nutrient concentration constant at the epithelium-stroma interface in the unperturbed steady state. We see that, similarly to what happened for model~4, decreasing $D_{\rm s}$ enhances the instability at long wavelengths which saturates as $D_{\rm s}$ goes toward zero. Increasing $D_{\rm s}$ instead tends to eliminate this maximum. In figure~\ref{Fig_Viscous_Nutrients}C, we study the effect of the diffusion coefficient $D$ in the epithelium compartment, following the same procedure. Here, increasing $D$ first favors the instability by increasing the thickness of the dividing region, but eventually suppresses the instability at short wavelengths. This is because increasing $D$ eventually renders cell division homogeneous over short lengthscales. Decreasing $D$ instead tends to stabilize the system, which stems from the fact that the thickness of the dividing region is decreased as nutrient penetration is decreased.

\begin{figure}[ht]
\scalebox{1.5}{
\includegraphics{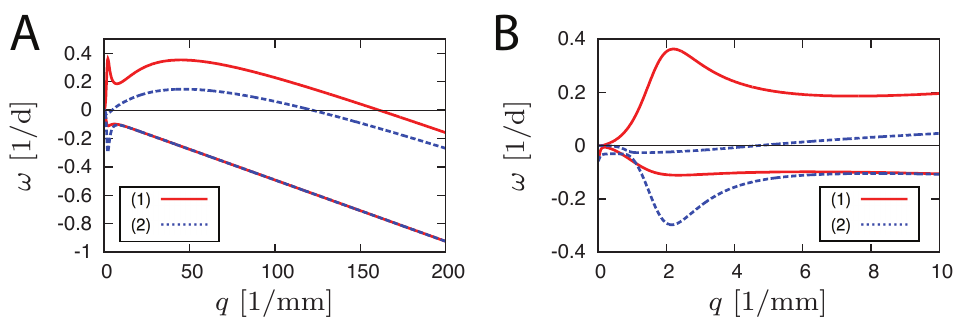}
}
\caption
{\label{Fig_Viscous_Nutrients_Comparison}
Comparison of the modes obtained with model~5 (linestyle (1)) and with model~2 (linestyle (2)), following the procedure described in the text. Parameters used for model~5 and their corresponding values in model~2 are identical to those of figure~\ref{Fig_Viscous_Nutrients}A. In model~2, the fitting parameters take the values $k=0.8$ division per day and $l=22$ $\mu$m. Panels (A) and (B) present the same mode structure but over two different wavevector ranges.
}
\end{figure}

As in the case of an elastic stoma, we compare in figure~\ref{Fig_Viscous_Nutrients_Comparison} the relaxation modes obtained with model~5 and those obtained with model~2 where the cell-production function is obtained by the fitting procedure described above.
As for figure~\ref{Fig_Elastic_Nutrients_Comparison}, we see that the instability is globally enhanced, and that especially at long wavelengths where a second maximum appears.

\subsection{Asymptotic behaviours of the modes}

In terms of the asymptotic expressions of the modes in the limit of small and large wave numbers, we find as for models~1 and 4 that the expressions associated with models~2 and 5 match to leading order in the short-wavelength limit and correspond to those given by equation~(20) of the Supplementary Information. Beyond that, the expressions are complicated and depend on nutrient coupling. We illustrate in equation~(35) of the Supplementary Information the asymptotic expression of one of these modes in a particular case to see how it differs from its equivalent for model~2 (equation~(21), Supplementary Information).

\newpage

\section{Discussion for a viscoelastic stroma: model~6}

We now consider the full model with coupling of the mechanical equations to the reaction-diffusion dynamics of nutrients, and with the full viscoelastic rheology for the stroma.

\subsection{Parameter-variation study}

\begin{figure}[ht]
\scalebox{0.73}{
\includegraphics{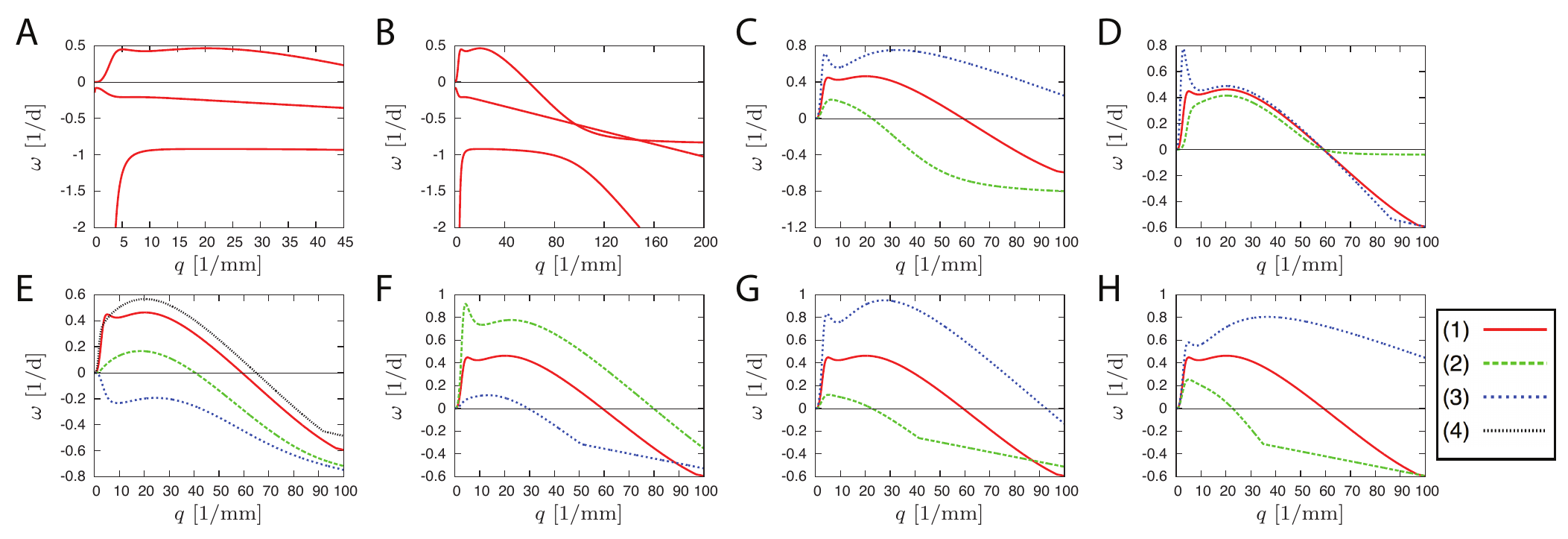}
}
\caption
{\label{Fig_ViscoEl_Nutrients}
Study of the relaxation-mode structure for model~6, as a function of several parameters entering the model. (A) and (B) The whole mode structure is shown for a default set of parameter values over two different wavector ranges. Parameter values are: $\eta=10$~MPa$\cdot$s, $\eta_{\rm s}=1$~MPa$\cdot$s, $\mu=10$ Pa, $\gamma_{\rm i}=5$~mN$\cdot$m$^{-1}$, $\gamma_{\rm a}=1$~mN$\cdot$m$^{-1}$, $\xi=10$~GPa$\cdot$s$\cdot$m$^{-1}$, $H=L=300$~$\mu$m, $d=75$~$\mu$m, $\kappa_1=100$~$10^{-6}$~m$^3\cdot$s$^{-1}$, $D=100$~$10^{-12}$~m$^2\cdot$s$^{-1}$, $D_{\rm s}=100$~$10^{-12}$~m$^2\cdot$s$^{-1}$, $c=0.1$~s$^{-1}$, $\bar{\rho}_0=0.6$~m$^{-3}$ and $v_{\rm off}=0$~m$\cdot$s$^{-1}$. Only three of the four modes are visible, since one relaxation rate takes very large negative values and stands out of scale. In each of the remaining panels, the most unstable mode is displayed for different parameter values, as one parameter at a time is varied as compared with the default parameter set of panels (A) and (B). The plain-red linestyle, associated with number (1), corresponds to the default parameter set. Other parameter values are as follows: (C) Variation of $\eta$: (2) $\eta=5$~MPa$\cdot$s ; (3) $\eta=20$~MPa$\cdot$s. (D) Variation of $\eta_{\rm s}$: (2) $\eta_{\rm s}=20$~MPa$\cdot$s, (3) $\eta_{\rm s}=100$~kPa$\cdot$s. (E) Variation of $H$: (2) $H=100$~$\mu$m ; (3) $H=50$~$\mu$m ; (4) $H=900$~$\mu$m. (F) Variation of $d$: (2) $d=50$~$\mu$m ; (3) $d=150$~$\mu$m. (G) Variation of $\kappa_1$: (2) $\kappa_1=50$~$10^{-6}$~m$^3\cdot$s$^{-1}$ ; (3) $\kappa_1=150$~$10^{-6}$~m$^3\cdot$s$^{-1}$. (H) Variation of $\gamma_{\rm i}$: (2) $\gamma_{\rm i}=10$~mN$\cdot$m$^{-1}$ ; (3) $\gamma_{\rm i}=2$~mN$\cdot$m$^{-1}$.
}
\end{figure}

In figure~\ref{Fig_ViscoEl_Nutrients}, we study the influence of some of the total 12 independent parameters on which the model depends. We choose here to present the dependence of the most unstable mode while independently varying six out of these parameters, which gives us an almost complete picture of the mode structure. The dependence in the diffusion constants $D$ and $D_{\rm s}$ has been studied in detail for models~4 and 5 and is therefore not reproduced here for model~6.
In figures~\ref{Fig_ViscoEl_Nutrients}A and \ref{Fig_ViscoEl_Nutrients}B, we show the mode structure for a default parameter set, which is then used in the other panels as a starting point for a systematic parameter-variation study. Only three out of the four modes are visible on the panels, since one relaxation rate takes very large absolute values and stands out of the plot. One of the four existing modes presents a region of instability. We note the presence of two distinct maxima in the unstable region, one around $q=5$ mm$^{-1}$ and one around $q=20$ mm$^{-1}$, corresponding to the two distinct maxima already discussed in the case of models~4 and 5. The viscoelastic relaxation rate $\tau^{-1}$ is equal to 0.86 per day here. We recognize features of the elastic model for the lower mode, which corresponds to relaxation rates larger than $\tau^{-1}$, and features of the viscous model for the two upper modes, which correspond to relatively long relaxation times. In figure~\ref{Fig_ViscoEl_Nutrients}B, we display the same mode structure as in figure~\ref{Fig_ViscoEl_Nutrients}A but over 
a larger wave-number domain, in order to visualize the convergence of the upper mode to $\tau^{-1}$ in the short-wavelength limit.

In the remaining panels, we study the dependence of the most unstable mode on different parameters. We first show the dependence on $\eta$, the epithelium viscosity in figure~\ref{Fig_ViscoEl_Nutrients}C. As expected and discussed in~\cite{Basan:2011fk}, the instability is increased with increasing values of $\eta$, which stems from the fact that the instability relies on viscous shear stresses in the epithelium. We recover this dominant tendency here. However, decreasing $\eta$ never really stabilizes the system as it was the case in~\cite{Basan:2011fk}, and the interface always remains unstable on a small interval at sufficiently small wave numbers. This can be attributed to the coupling to nutrient diffusion, similarly to what happens in the context of crystal growth, where the system is always unstable at arbitrarily long wavelengths~\cite{mullins1964stability,langer1980instabilities}. In figure~\ref{Fig_ViscoEl_Nutrients}D, we study the dependence on the stroma viscosity $\eta_{\rm s}$. Increasing $\eta_{\rm s}$ to very large values tends to eliminate the low-$q$ maximum, but is not enough to render the system stable. It is interesting to note that, contrary to the case of the Saffman-Taylor instability~\cite{saffman1958penetration,bensimon1986viscous}, which occurs when a fluid of lower viscosity displaces a more viscous one in a Hele-Shaw cell, the relative values of the two fluid viscosities here is not a criterion to change the stability of the system. This fact was also observed in the context of model~2 and illustrated in figure~2 of the Supplementary Information, section~3. In figure~\ref{Fig_ViscoEl_Nutrients}E, we study the dependence in $H$, the thickness of the epithelium layer. A sufficiently thin epithelium does not display any instability. This is because the underlying mechanism of the instability remains the presence of viscous shear stresses within the epithelium, even when nutrient diffusion is introduced. In figure~\ref{Fig_ViscoEl_Nutrients}F, we then study the dependence on $d$, the distance between the epithelium-stroma interface and the source of nutrients in the stroma. Varying this distance has a pronounced effect on the growth rate of the instability. This is natural since this distance directly influences the distribution of nutrients within the stroma, and so dramatically influences the stability of the system. In figure~\ref{Fig_ViscoEl_Nutrients}G, we study the dependence on $\kappa_1$, which describes the coupling of cell division to nutrient concentration (equation~(\ref{CellProdFctNutrients})). Decreasing $\kappa_1$ allows only the long-wavelength maximum to remain unstable, and increasing it allows the other maximum to reach larger values. Finally, figure~\ref{Fig_ViscoEl_Nutrients}H shows that the epithelium-stroma interfacial tension $\gamma_{\rm i}$ has a stabilizing effect, and that primarily at short wavelengths.

Among the parameters whose variations have not been displayed in this figure, $D$ and $D_{\rm s}$ have similar effects to those seen in figures~\ref{Fig_Elastic_Nutrients} and \ref{Fig_Viscous_Nutrients} in the frameworks of models~4 and 5, and the remaining parameters have a simple, intuitive effect: Increasing the elastic modulus $\mu$ of the stroma tends to stabilize the system, and so does the increase of $c$, the rate of nutrient consumption by the epithelial cells. On the other hand, the stroma thickness $L$ as well as the total amount of nutrient produced $\bar{\rho}_0$ have a destabilizing effect (graphs not shown).

\subsection{Mode structure and asymptotic expressions}

Similarly as before, the different associated expressions of the fast modes to leading order are unchanged as compared with the case of model~3. In the limit of small wavelengths, we retrieve the expressions given by equation~(\ref{largeqModesViscoElSimple}) to linear order, which is explained for the three first modes by the fact that these correspond to fast relaxations of the system, and for the fourth mode because its limit corresponds to the inverse relaxation time of the viscoelastic material constituting the stroma, which is independent of cell division. In the long-wavelength regime, the expressions to leading order of the two first modes are identical to those obtained without nutrient diffusion and are given by equation~(\ref{smallqModesViscoElSimple}). The situation is different for the modes of order constant and proportional to $q^4$, for which the specific coefficients do depend on nutrient coupling. The general expressions are very long, but the example of equation~(35) of the Supplementary Information discussed above holds also here.

\section{Discussion}

In this work, we have proposed a mechanism for the formation of undulations at the epithelium-stroma interface that arise from a physical instability intrinsic to the structure of a multilayered epithelium. This instability, presented originally in~\cite{Basan:2011fk} and discussed here in more details, is a hydrodynamic instability arising from viscous shear stress originating from flow within the tissue due to cell renewal. Such an instability can develop in a healthy epithelium, depending on its physical characteristics such as the thickness of the cell-division region, its long-term viscosity due to cell-cell adhesion, and the mechanical resistance of the stroma. These results are in agreement with the observation that the degree of undulations {\it in-vivo} typically depends on the grade of the dysplasic epithelium. The grade itself correlates with the number of dividing cell layers~\cite{bouquot2006epithelial}, and is part of the diagnosis of dysplasic tissues, for example of oral cancer~\cite{Tsai:2008ys,Hamdoon:2012vn}. The mechanical resistance of the stromal tissue can also influence the morphology of the basement membrane. This has been observed for example to a mild extend {\it in-vivo} in the stratefied epithelia of the cornea and of the limbus~\cite{Grueterich:2003kx}, and in a clearer way in reconstituted epithelia {\it in-vitro}~\cite{Barbaro:2009fk}. There, different degrees of undulations were observed as a function of the properties of the supporting scaffold on which the tissue was grown. More generally, this instability could be present in all sufficiently viscous fluids with internal flow due to material production or destruction. Growth factors, proteases, external feedback and abnormal proliferation are not required for the occurrence of this instability.

Nevertheless, introducing coupling of cell division to the diffusion of nutrients, growth factors or oxygen from the stroma adds a destabilizing effect, similar to that occurring during diffusion-limited growth~\cite{Langer:1989fk,ben1990formation}. This leads to a significant enhancement of the previously proposed instability and to the appearance of an additional maximum of the unstable mode at long wavelengths. Instabilities originating from similar coupling terms have been identified in living systems, for example in the case of bacterial-colony growths, where similar patterns to those associated with aggregation phenomena and viscous fingering have been observed~\cite{matsushita1990diffusion,Ben-Jacob:1994fk,ben2000cooperative}. Such a coupling can lead to fractal branching patterns via the process of diffusion-limited aggregation~\cite{sander1986fractal,fujikawa1989fractal} or other types of branching patterns depending on the bacterial morphotype~\cite{ben1994holotransformations,Ben-Jacob:1995fk}.

The destabilizing effect of the coupling to nutrient diffusion is reminiscent of the Mullins-Sekerka instability in the context of crystal growth~\cite{mullins1964stability,langer1980instabilities}. This analogy relies on the fact that nutrients here play a similar role as either latent heat when crystal growth occurs in an undercooled melt of a pure substance or solute molecules when it occurs in an isothermal but binary, supersaturated solution: outward pointing epithelium protrusions have access to a larger amount of nutrients than retracted regions, therefore proliferate faster, enhancing the already existing protrusion. The instability therefore develops faster when nutrient diffusion is slower, especially at long wavelengths as illustrated in figures~\ref{Fig_Elastic_Nutrients} and \ref{Fig_Viscous_Nutrients}. Nevertheless, differences exist between the diffusion-driven instability proposed here and that of growing crystals. For one, in the latter case, the solid phase grows by addition of new material coming from the environment to the interface. This phase therefore grows only at the surface and, once formed, remains rigid and static in the bulk. Here however, the new material comes from cell division within the epithelial tissue itself, and growth occurs as a bulk phenomenon. As a consequence, the instability proposed here can only develop for sufficiently thick epithelia, as illustrated in figure \ref{Fig_ViscoEl_Nutrients}E, and its driving force still relies on differential cell flows, as proposed originally~\cite{Basan:2011fk}.

The instability discussed in this work may also evoke the Saffman-Taylor instability, which occurs when a fluid of lower viscosity displaces a more viscous one in a Hele-Shaw cell~\cite{saffman1958penetration,bensimon1986viscous}. However, the two mechanisms are very different and should not be confounded. In the case of the Saffman-Taylor instability, an external force in the form of a global pressure difference is required to drive the system out of equilibrium. The dynamics is governed by Darcy's law arising from the balance of pressure gradient with friction of the fluid with the background. The Laplace equation for the pressure field together with the constant-pressure boundary conditions determine the instability and the mode structure. On the other hand, in the case studied here, forces are generated within the epithelial tissue by internal processes. In addition, the field responsible for the second, long-wavelength maximum is the nutrient-concentration field, which follows a Laplace equation in the stromal compartment in the regime of fast diffusion, rather than the pressure field. A particular illustration of the difference between the two instabilities can be seen from figure~\ref{Fig_ViscoEl_Nutrients} of the main text and figure~2 of the Supplementary Information, which show that here, the instability can develop when either of the two fluids---epithelium or stroma ---is the more viscous one.

Beyond the mechanisms discussed in this work, which rely entirely on cell-division driven cellular flows and their coupling to passive diffusion, there exist more specific, active, biological mechanisms that can cause significant aberrations of the tissue interface in malignant cases. As one example, neoplastic epithelial cells are known to excrete signaling molecules that recruit and stimulate stromal fibroblasts and induce the differentiation of monocytes into macrophages~\cite{weinberg2007bc,Hanahan:2011ly}. This further induces the production and release of epithelial growth factors, matrix metalloproteinases and angiogenetic factors~\cite{weinberg2007bc,Kalluri:2006ys,Qian:2010uq}. This effect may enhance the progression of protrusions in a positive feedback loop, since advancing protrusions increase the concentration of signaling molecules in regions of the stroma close to them, further stimulating the expression of growth factors by fibroblasts. As a second example, many malignant epithelial tissues and their stromal-associated cells are known to secrete matrix-degrading enzymes such as metalloproteinases~\cite{weinberg2007bc,Friedl:2008uq,Kessenbrock:2010ly,Hanahan:2011ly}. These enzymes play a central role for the invasion of epithelial protrusions in the stroma by cutting through the dense meshwork of collagen fibers, carving out space for the expanding tumour. Such proteases like the matrix metalloproteinases~2 and 9 can be produced by the tumour-associated macrophages~\cite{Kessenbrock:2010ly,Qian:2010uq}, a mechanism that contributes to the cross-talk mentioned above.

While neither of these effects is directly included in our model, we expect the driving force behind the advancing protrusions to still originate from the fundamental mechanical picture proposed here. Indeed, these effects may enhance an already existing instability, but do not represent an alternative driving force to the present model. The situation is fundamentally different in the case, where cells from the epithelium have undergone the transition to a motile phenotype and invade the stroma~\cite{Friedl:2009fk,Friedl:2012fk}. Though clearly only relevant for malignant tissues, this mechanism is fundamentally different from the one discussed in this work and provides a true alternative process that can lead to pronounced irregular pathologies observed {\it in vivo}~\cite{tavassoli2003pag}, and ultimately drive cancerous invasion. However, one may speculate that the acquisition of invasive motile phenotypes itself comes after an original protrusion has formed rather than be responsible for its initiation. In two-dimensional tissue migration for example, the emergence of fingering protrusions seems to arise spontaneously~\cite{mark2010physical} and is often accompanied by the emergence of a different phenotype at the tip of the protrusion~\cite{Poujade:2007fk}. The appearance of such leader cells may however be the signature of a loss of contact inhibition~\cite{Drees:2005qq,Basan:2010kx,Puliafito:2012fk}, and may therefore be induced by the original perturbation rather than be responsible for its initiation. One may therefore speculate that a physical mechanism such as the instability proposed here could initiate the emergence of invasive phenotypes.

\ack
We thank Profs. J.~Prost and J.-F.~Joanny for thinking about the original ideas with us as well as for numerous constructive discussions. The authors acknowledge support from the European network MitoSys.

\end{document}